\documentclass[authoryear]{elsarticle}

\usepackage{natbib}

\usepackage{graphics}
\usepackage{graphicx}
\usepackage{epsfig}
\usepackage{color}           

\usepackage{amssymb}

\renewcommand{\vec}[1]{ {\mathbf #1} }

\newcommand{\xx}{ \vec x}

\newcommand{\aap}{{\it Astron. Astrophys. }}

\newcommand{\apj}{{\it Astrophys. J. }}
\newcommand{\apjl}{{\it Astrophys. J. Lett. }}

\newcommand{\nat}{{\it Nature }}

\newcommand{\solphys}{{\it Solar Phys. }}

\begin{document}

\begin{frontmatter}



\title{Simulation of the Observed Coronal Kink Instability and Its Implications for the SDO/AIA}


\author{A.K. Srivastava$^a$, G.J.J. Botha$^b$, T.D. Arber$^b$, P. Kayshap$^{a}$}

\address{$^a$Aryabhatta Research Institute of Observational Sciences (ARIES), Manora Peak, Nainital-263 129, India.}
\ead{aks@aries.res.in}
\address{$^b$Physics Department, Centre for Fusion, Space and Astrophysics, University of Warwick, Coventry CV4 7AL, UK.}

\begin{abstract}
\cite{SrivEA10} have observed a highly twisted coronal loop, 
which was anchored in AR10960 during the period 04:43 UT-04:52 UT
on 4 June 2007. The loop length and radius are approximately 80 Mm
and 4 Mm, with a twist of 11.5 $\pi$. These observations are used
as initial conditions in a three dimensional nonlinear magnetohydrodynamic 
simulation with parallel thermal conduction included. The initial 
unstable equilibrium evolves into the kink instability, from
which synthetic observables are generated for various 
high-temperature filters of SDO/AIA. These observables include
temporal and spatial averaging to account for the 
resolution and exposure times of SDO/AIA images. Using the simulation
results, we describe the implications of coronal kink instability 
as observables in SDO/AIA filters. 
\end{abstract}

\begin{keyword}
magnetohydrodynamics (MHD) \sep magnetic reconnection \sep corona
\end{keyword}

\end{frontmatter}

\clearpage
\section{Introduction}

The active region magnetic field exhibits complex topology
{\bf due to the emergengce of new large-scale magnetic flux}
from the 
sub-photospheric layers of the Sun.
Such complexity some times generates various types of 
magnetic instabilities that are well observed in the solar active 
regions as a trigger of flares and associated dynamical processes
\citep{SrivEA10,Ku10,Claire11,Innes12}. Among the various types
of magnetic instabilities, the kink instability is most commonly 
observed in the solar atmosphere, and evolved in the 
large-scale magnetic flux-tubes due to their azimuthal twist. 
However, the twist must cross the minimum threshold 
value of $\Phi >2.5 \pi$ to trigger the kink instability
in the magnetic flux-tubes in the solar corona \citep{Alan79}.
The kink instability is well observed in active region loops
\citep{SrivEA10}, eruptive filaments \citep{Liu08}, 
eruptive coronal cavities \citep{Liu07}, and
it can eventually trigger flares and CMEs \citep{Kleim06,Cho09}.
Recently, \cite{Zaqa10} have given an important 
conclusion that the axial mass flow in the magnetic flux-tubes reduces 
the threshold of kink instability, and can easily lead to 
solar flares and CMEs. The kink instability does not only 
liberate the flare energy, it can also cause the 
excitation of MHD wave modes \citep{Hy08}.
The kink instabilities observed in the astrophysical plasma 
are also firstly reproduced in the laboratory experiment
\citep{Moser12}, which shows the confinement of kink unstable 
twisted plasma structures due to multi-stage cascade reconnection
at the laboratory scales.

In addition to the observations the coronal kink
instability is studied using analytical and numerical modelling
\citep{MikicEA90,BatyHeyvaerts96,ArberEA99,GerrardEA02}. 
The ideal MHD kink instability has been inferred as 
a trigger for reconnection in the flaring loops 
\citep{BrowningEA2008,HoodEA09}. 
\cite{BothaEA11} initialised the kink instability with a twisted magnetic field  
and reported that the inclusion of thermal conduction along 
magnetic field lines can reduce the maximum reconnection generated temperature 
by an order of magnitude. This changes the range of spectral lines as observables
of the kink instability when compared to simulations without conduction. 
\cite{BothaEA12} have also reported the observational signatures of the MHD kink instability 
with the inclusion of thermal conduction. A kink-unstable cylindrical loop 
is evolved by using the initial conditions of a highly twisted loop as observed by TRACE \citep{SrivEA10}.
The numerical results are synthesized through TRACE temperature response 
function filtering, and through spatial and temporal
averaged line of sight intensity measurements
\citep{HaynesArber07}.

In the present paper, we extend the work of \cite{BothaEA12} and study the 
evolution of the kink unstable loop as observed by \cite{SrivEA10} in the 
high-temperature filters of SDO/AIA. 
In Sec.~2 we present a brief description of the numerical model and 
initial conditions. In Sec.~3 we describe the results related to the
synthesis of a kink unstable loop in various SDO/AIA channels. 
The discussion and conclusions form the last section.

\section{Numerical Simulation of the Observed Coronal Kink Instability}

The nonlinear three-dimensional simulation is executed by implementing the MHD 
Lagrangian-remap code (Lare3d) of \cite{ArberEA01}.
The numerical code solves the resistive MHD equations for the fully ionised plasma with a heat 
flux included in the energy equation \citep{BothaEA11, BothaEA12}. 
Thermal conduction is considered to be included along the 
magnetic field lines in the form of a classical \cite{SpitzerHarm53} or 
Braginskii conductivity with $\log\Lambda=18.4$ that corresponds to 
the standard thermal conductivity parallel to the magnetic field of 
$\kappa_\parallel =10^{-11}\,T^{5/2}$ W m$^{-1}$ K$^{-1}$  
\citep{Priest00, BothaEA12}.
The numerical code consists of an artificial resistivity that is activated only when 
the {\bf electric} current exceeds a critical value, which is set in the present case 
{\bf to} $j_c=2$ mA \citep{BothaEA12}. This resistivity is considered in the form given as follows
\begin{equation}
\eta = \left\{ \begin{array}{ll}
                 \eta_0, & \quad |j|\geq j_c, \\
                  0,      & \quad|j|< j_c,
               \end{array} 
        \right.
\end{equation}
where $\eta_0$ is the anomalous resistivity. 
$\eta_0$ is switched on when the kink instability occurs 
and thereafter it remains active for the duration of the simulation 
\citep{BothaEA11, BothaEA12}. The coronal loop is initialised as a straight twisted cylinder in a 
uniform background temperature and density (Fig.~1) as previously performed 
by \cite{BothaEA12}. 

The initial conditions of the numerical simulation of the coronal loop is taken 
from observations of a highly twisted loop \citep{SrivEA10}. 
Using the multi-wavelength observations of SOHO/MDI, SOT-Hinode/blue-continuum (4504 \AA ), G band (4305 \AA ), 
Ca II H (3968 \AA ), and TRACE 171 \AA , 
they observed a highly twisted magnetic loop in AR 10960 during the period 04:43 UT-04:52 UT on 2007 June 4. 
SOT-Hinode/blue-continuum (4504 \AA ) observations show that the penumbral filaments of a positive polarity sunspot 
have counterclockwise twist that may be caused by the clockwise rotation of the spot umbrae and can activate
a right-handed helical twist in the loop system. This loop whose one footpoint is anchored in this sunspot, 
shows strong right-handed twist in chromospheric SOT-Hinode/Ca II H (3968 \AA ) and coronal TRACE 171 \AA~images, which
was consistent with the Hemisphere Helicity Rule (HHR). The length and the radius of the loop are estimated 
as L$\approx$80 Mm and a$\approx$4.0 Mm, respectively. 
The distance between neighbouring turns of magnetic field lines is estimated as $\approx$10 Mm, which 
further gives the total twist angle $\Phi$=11.5$\pi$ as estimated for a homogeneous distribution of the twist 
along the loop. This observed twist
is much larger than the Kruskal-Shafranov instability criterion {\bf ($\Phi$$>$2$\pi$)} for the kink instability and 
later triggers  a B5.0 class solar flare that occurred between 04:40 UT and 04:51 UT in this active region.
The details of these observational finding are given in \cite{SrivEA10}.

\clearpage

\begin{figure}
\centerline{
\mbox{
\includegraphics[width=4.0cm]{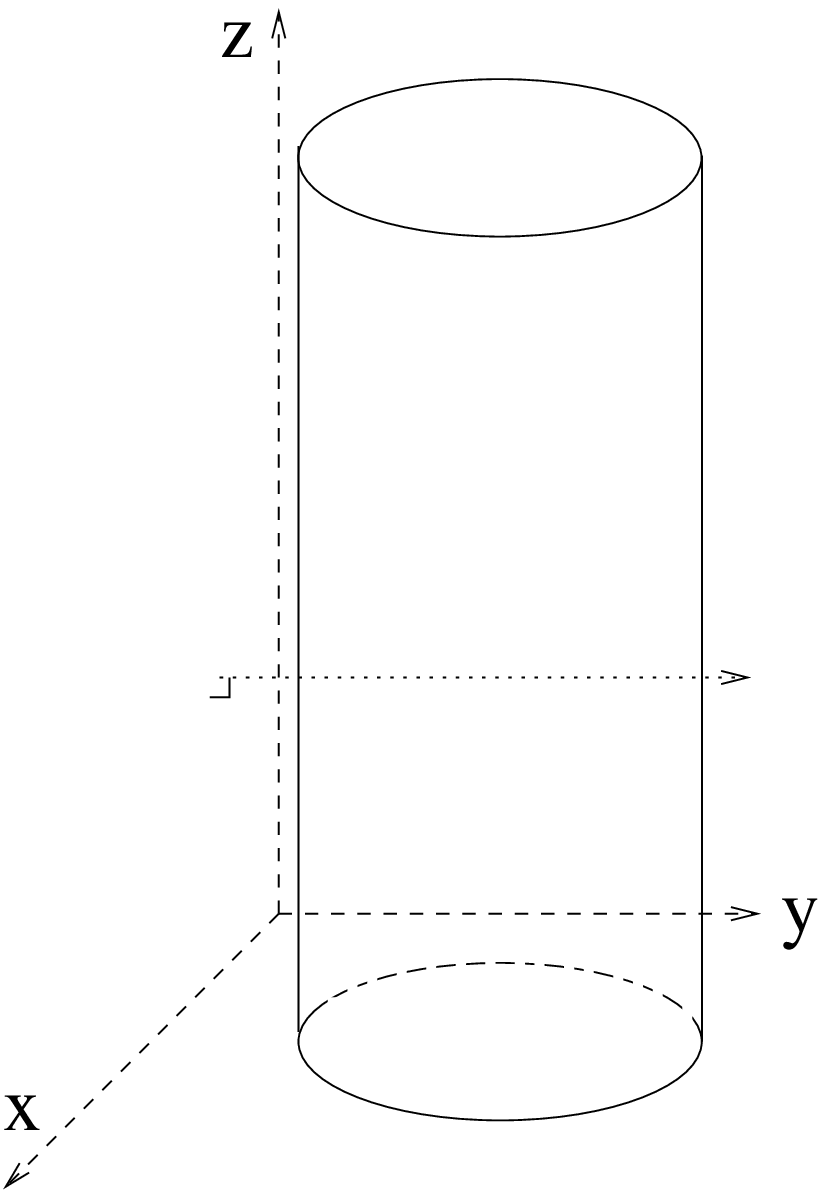}
\includegraphics[width=6cm]{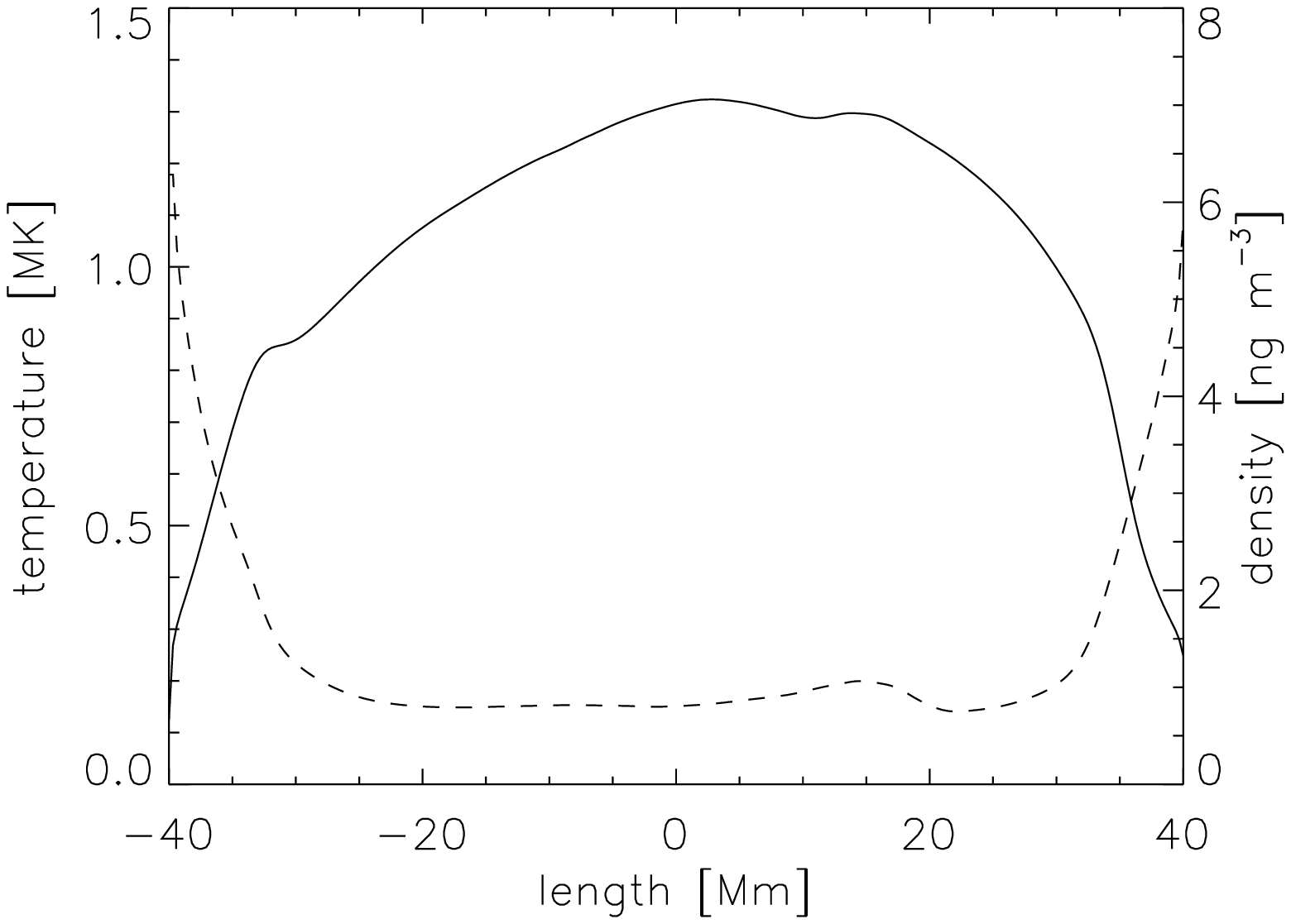}}}
\caption{Left-panel: The cartoon shows an orientation of the cylindrical loop in the 
         Cartesian numerical domain. The dotted line indicates about the 
         integration path of the line of sight integral as described by Eq.3 
         along the $y$-direction and perpendicular to the $x-z$ plane.
         Right-panel: Temperature and mass density profiles along the length of the cylindrical loop on 
         its central axis  at a time of 579.98 s.
         The solid (dashed) line is the temperature (density). 
         Initialisation is occurred with a constant temperature of 0.125 MK and a 
         constant density of 1.67 ng m$^{-3}$.
         }
\label{fig:tempdens}
\end{figure}

\begin{figure}
\centerline{
\mbox{
\includegraphics[width=7cm]{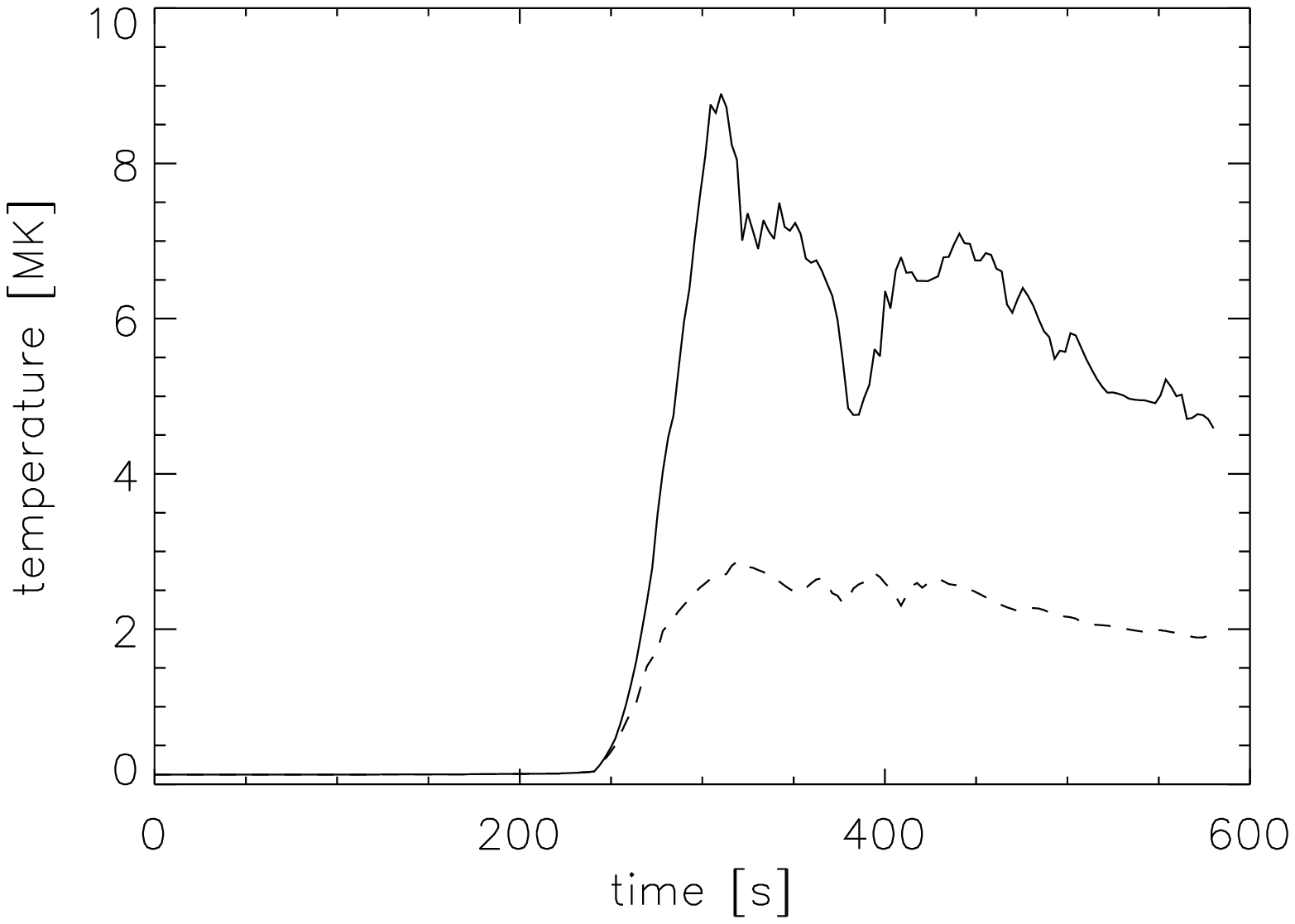}}
\includegraphics[width=7cm]{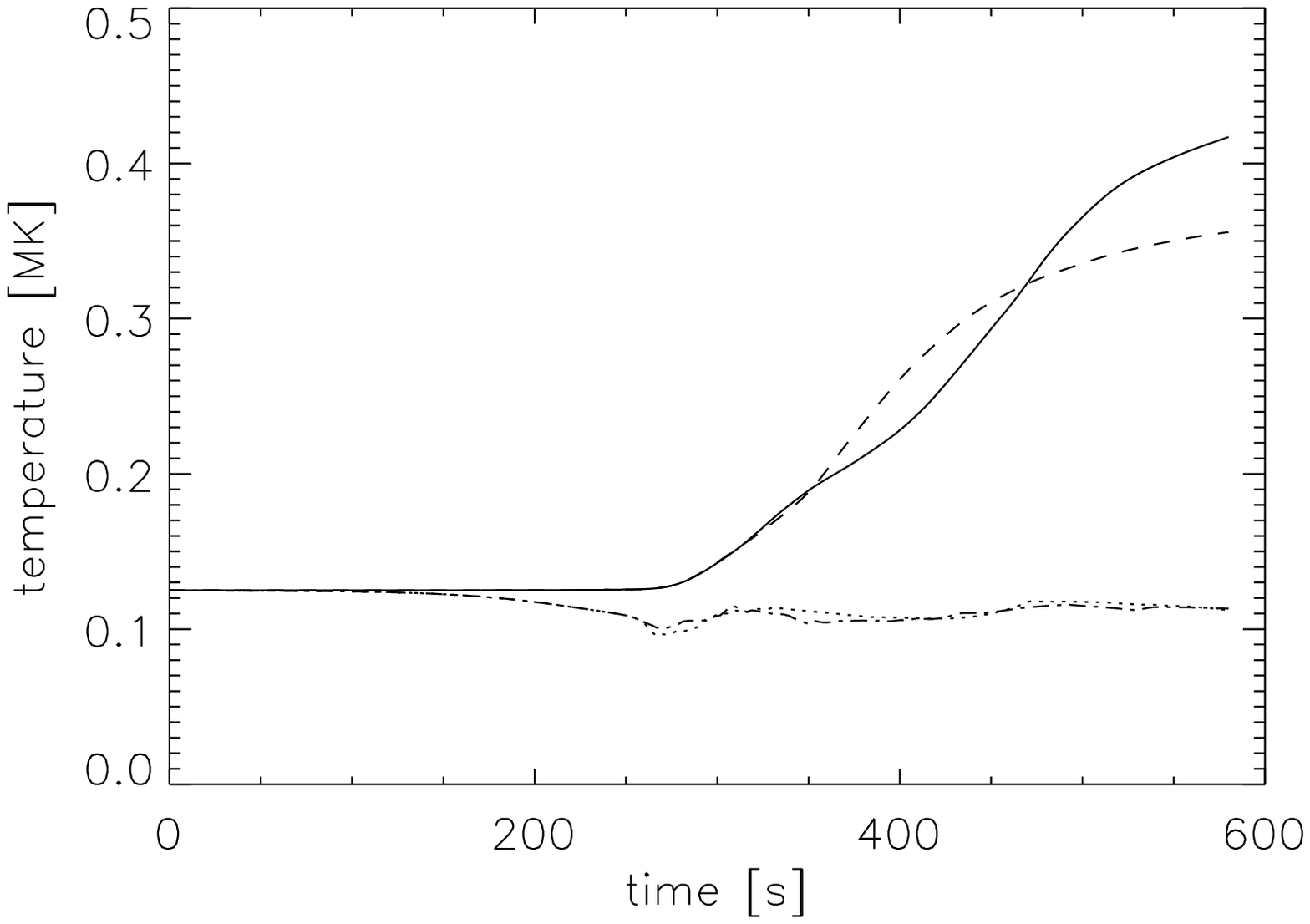}}
\caption{Left-Panel: Maximum temperature during the evolution of the kink instability in which the solid (broken) line
is without (with) thermal conductivity. Right-Panel : The average temperature
during the evolution of the kink instability in which the solid (broken) line is without (with)
thermal conductivity. The minimum temperature is also represented  
by the dash-dot-dashed (dotted) line without (with) thermal conduction. 
         }
\label{fig:kink}
\end{figure}

{\bf The observed curved loop is straightened to perform  numerical simulations (Figure 1, left-panel).}
One of the footpoints of the active region loop is anchored in a positive 
polarity spot \citep{SrivEA10} and it is estimated that the minimum value of the  
magnetic field at this location is 470 G \citep{BothaEA12}. From this photospheric value of the magnetic field, 
the chromospheric field is calculated approximately as 80 G, as well as near the loop
apex in the corona as 20 G \citep{PetPat09, BothaEA12}. 
The numerical simulation is initialised for the loop with a maximum internal field 
of 20 G, while the outside background magnetic field is considered as a
uniform value of 15 G  parallel to the cylindrical axis of the simulated loop structure. 
In the numerical simulations the coronal loop is initialised as a uniform cylinder 
in force-free equilibrium and set unstable to an ideal MHD kink instability \citep{HoodEA09, BothaEA12}.
The axial twist is given by the following equation
\begin{equation}
\Phi = \frac{LB_\theta}{rB_z} \quad\mbox{with}\quad \max(\Phi)=11.5\pi,
\label{eq:twist}
\end{equation}
where $L$ is the loop length and $r$ is the radial distance from the central axis of the loop. 
The axial and azimuthal magnetic fields are given respectively by $B_z$ and $B_\theta$
as a function of $r$. The 
maximum twist is considered at position $r=1$ Mm, which 
exceeds the stability threshold ,therefore, the loop becomes
kink unstable \citep{MikicEA90}. The details of the magnetic field 
configuration is outlined in \cite{BothaEA11, BothaEA12}. 
Gravity is not included in the simulations. 
The initial temperature and mass density are taken as uniform and constant, with a typical  
values of $1.67\times10^{-12}$ kg m$^{-3}$ \citep{YoungEA09} and 0.125 MK respectively. 
This temperature is selected to make the evolution 
of the kink instability as evident in the SDO/AIA filters.  
During the evolution of the kink instability the temperature increases locally 
where reconnection occurs in the simulated loop \citep{BothaEA11}. This temperature increament is sufficient
that the results are highly insensitive to the lower initial temperatures.
The initial temperature was chosen specifically so that these high temperatures 
(Figs.~\ref{fig:tempdens}, right-panel and \ref{fig:kink}) 
will be visible using the SDO/AIA filter response functions (Fig.~\ref{fig:AIAresponses}).
The details of these technical aspects of the numerical simulation are described in \cite{BothaEA12}.

\section{Synthesized Kink Unstable Loop as an Observable in Various Filters of SDO/AIA}

The evolution of the kink unstable coronal loop is studied in the Cartesian geometry 
(Fig.~1)  of the simulation domain. 
The boundaries in the $x-y$ plane perpendicular to the loop axis are considered at 
$\pm 8$ Mm and it is set to be reflective. Therefore, there are 4 Mm seperation between the loop edge and 
the outer boundaries to avoid the numerical reflections
near the boundaries. 
Along the loop axis, these boundaries are cosidered at $\pm 40$ Mm with velocities and 
temperature fixed respectively at zero and the initial background value by allowing 
the temperature gradients. Therefore, the heat flux across the ends of the loops exists
in the simulation domain. 
The grid resolution in $x-y-z$ is given by $128\times 128\times 256$.
These technical details are summarized in \cite{BothaEA12}.

\begin{figure*}
\centerline{
\includegraphics[width=15cm]{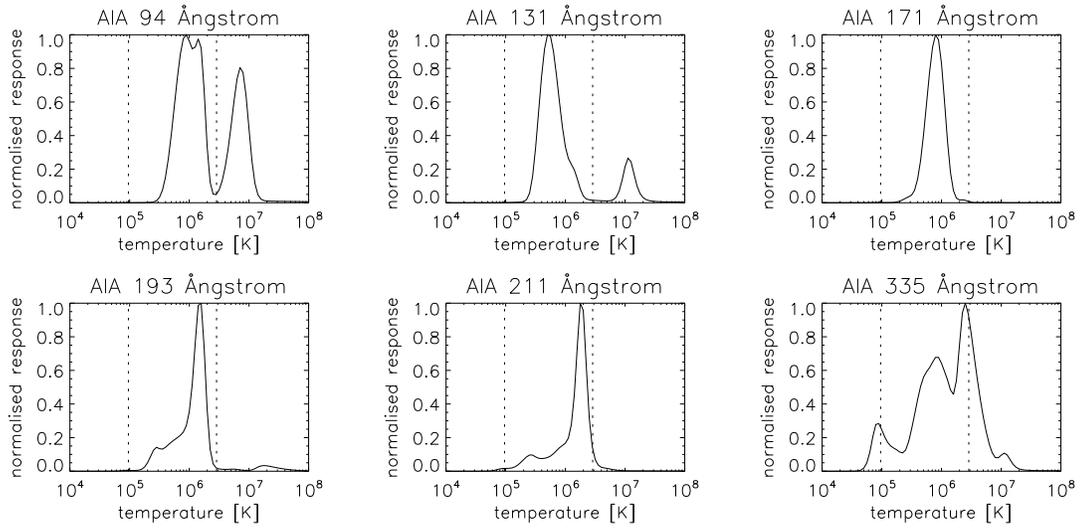}}
\caption{Normalised SDO/AIA temperature response functions with the minimum and maximum 
         temperatures during the simulation represented by vertical dotted lines. 
         The 304~\AA~band captures chromospheric and transition region temperatures, 
         features that are not included in our model. Consequently it is 
         not included with the other filters.
         }
\label{fig:AIAresponses}
\end{figure*}

The images generated from the numerical simulations were obtained by using the 
modified temperature response functions  
of SDO/AIA \citep{AshBoe11}. Details of the line contributions for 
the AIA channels are outlined in \cite{ODwyerEA10} also. 
The emission is calculated at every node of the numerical grid and thereafter integrated 
along the $y$-direction perpendicular to the axis of the simulated loop (cf., Fig.~\ref{fig:tempdens}, 
left-panel). The line of sight intensity integral is given by  
\begin{equation}
I = \int^{+L_y}_{-L_y} g(T)\rho^2 dy
\label{eq:line}
\end{equation}
where $I$, $g(T)$, $\rho$ are respectively the measured intensity, the temperature response function of 
the various AIA filters, and the mass density. Fig.~\ref{fig:AIAresponses} 
presents the temperature response functions of all AIA filters. This 
integration produces intensity images in the $x-z$ plane, which are then 
integrated over time with the typical time interval determined by the exposure time for 
SDO/AIA as 2.9 s. After this procedure, the time integrated image is degraded by spatially 
averaging over squares of $0.375\times 0.375$ Mm$^2$ to compensate for the pixel 
resolution of SDO/AIA. The similar process is adopted for the generation of synthesised 
TRACE observables as reported by \cite{BothaEA12}. However, in the present study,
we perform this image synthesis of the kink unstable loop as an observable in various filters of SDO/AIA
(cf., Figs. 4-6). {\bf It should be noted that we used corrected AIA 94 response function with
correction for enhanced response at log(T)$<$6.2 due to missing Fe IX, Fe XII lines
in CHIANTI. Similarly, the response function for AIA 131 \AA~ is used with
considering the contribution of Fe VIII and Fe XI. This may vary overall intensity features 
in the synthesized images of 094 \AA\ and 131 \AA\ compared to those generated by using old response functions.
The 94 \AA~, 211 \AA~ and 131 \AA~ images are synthesized by choosing the elimination of lowest 1\% of the intensity values at time 321.9 s.}

After initialisation, the simulated kink unstable loop evolves through a linear phase with a duration of
300 s (Fig.~\ref{fig:kink}).  During the nonlinear 
phase, the kink instability drives magnetic field into a current sheet where 
reconnection occurs and the temperature reaches to a maximum value due to the released energy
and subsequent heating. 
Thereafter, thermal conduction along the magnetic field lines transports the generated heat flux 
along the field lines away from the reconnection sites where the plasma was 
heated. The time-scale for the evolution of the kink 
instability in all SDO/AIA filters is the same as of the case without thermal conduction. 
The physical processes 
during the kink instability, with and without thermal conduction, are discussed 
in comprehensive manner by \cite{BothaEA11} and \cite{BothaEA12}. 
In the present numerical simulations, the thermal conductivity causes 
the spread of temperature along the magnetic field lines resulting in the synthesized images in which the 
features are less resolved.

The synthesized images 
show that the footpoints of the loops increase their emission during the nonlinear phase 
of the kink instability irrespective of the inclusion or exclusion of thermal conductivity. The line of sight intensity 
integral (Eq.3) is determined by the 
temperature response function of SDO/AIA and the chosen mass density. Fig.~\ref{fig:tempdens} (right-panel) 
gives the temperature and mass density profiles for the 
numerical simulation including thermal conduction. It is noticiable that the enhanced emission is due to 
a density increase at the footpoints of the simulated loop and it is not due to the footpoint heating. Plasma  
flows from the middle of the numerical domain where the current sheets are formed as reconnection sites 
, driven by MHD ponderomotive forces generated during the evolution of the kink instability. 
Footpoint brightening due to compression were also evident in the coronal loop simulations 
of \cite{HaynesArber07} and \cite{BothaEA12}. 
The factor that determines the initiation of the nonlinear phase of the kink instability is 
the initial twist (Eq.2) in the coronal loop. A smaller value of maximum twist  
increases the duration of the linear phase and vice versa. However, it is found that once the nonlinear 
phase is evolved, the formation of the current sheet, the reconnection, and the thermal 
aftermath have the same duration as long as the initial twist exceeds the stability 
threshold \citep{BothaEA12}. These physical and technical details are discussed in detail 
by \cite{BothaEA12}. 

\begin{figure*}
\centerline{
\includegraphics[width=15cm]{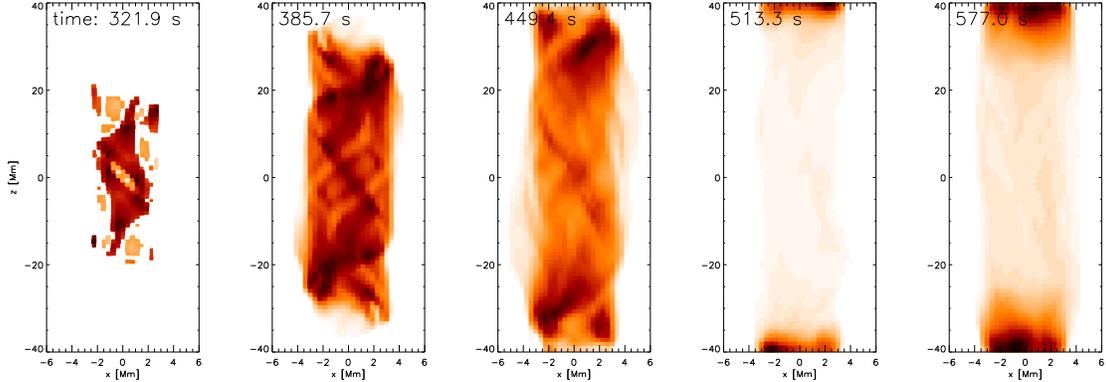}}
\caption{AIA 94 \AA~{\bf simulations} with exposure time 2.9 s and spatial resolution 
         0.375 Mm per pixel. The intensity is in reverse colour, with white the lowest 
         value on the scale. This minimum is chosen so that the lowest 1\% of the values 
         at time 321.9 s are eliminated from these plots. All the exposures use the same 
         reverse colour scale.
         }
\label{fig:AIA94}
\end{figure*}

During the study of the evolution of such kink instability and its comparison with the observations, it is more realistic to compare 
the obtained results when thermal conduction is included in the numerical simulation model. 
The first snapshot in Fig.~\ref{fig:AIA94} is synthesized at 321.9 s for AIA 94 \AA~and 
shows the twisted structure of the current sheet formed at the initiation of the 
nonlinear phase of the evolution of kink instability. 
In the later images, the thermal conduction has transported the reconnection generated heat flux from 
the internal reconnection sites along the magnetic field lines. 
As the internal reconnection occurs within the simulated loop, the magnetic 
field lines become straight along the length of the loop \citep{HaynesArber07, BothaEA12}. 
At the same time, the heat flux is transported along magnetic field lines away from the reconnection 
sites. Therefore, the loop becomes cool and its internal magneto-plasma structures are 
less resolved \citep{BothaEA12}. 

The corrected and modified temperature response functions for the various wavebands (filters) of SDO/AIA are presented in 
Fig.~\ref{fig:AIAresponses}.
The emissions in 171 \AA~for SDO/AIA and TRACE are almost the same.
The only difference between 
these two observational instruments is the time resolution. The TRACE exposure time 
is approximately eleven times longer than that of the SDO/AIA. Therefore, 
the images from SDO/AIA are much sharper. 
\cite{BothaEA12} have already reported the kink unstable loop as an observable in 
TRACE 171 \AA\ filter, therefore, we do not present the similar results as obtained 
in AIA 171 \AA\ filter. In the present paper, we report the evolution of 
observed kink unstable loop as an observable in the high-temperature filters of
SDO/AIA. 
Moreover, AIA 193 \AA\ and 211 \AA\ filter responses are almost similar, therefore,
we present only the results related to the 211 \AA\ synthesized AIA images.
The 304~\AA~band captures chromospheric and transition region temperatures 
and related features that are not included in our model. 

\begin{figure*}
\centerline{
\includegraphics[width=15cm]{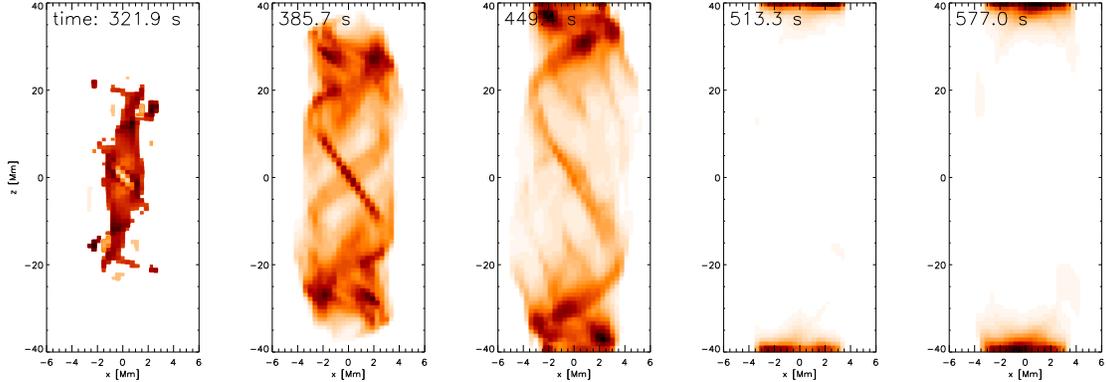}}
\caption{AIA 131 \AA~simulations with exposure time 2.9 s and spatial resolution 
         0.375 Mm per pixel. The intensity is in reverse colour, with white the lowest 
         value on the scale. This minimum is chosen so that the lowest 1\% of the values 
         at time 321.9 s are eliminated from these plots. All the exposures use the same 
         reverse colour scale.
         }
\label{fig:AIA131}
\end{figure*}

The emission images for SDO/AIA 94 \AA, 131 \AA~and 211 \AA~are presented in 
Figs. \ref{fig:AIA94}-\ref{fig:AIA211}. As each temperature response function 
(Fig.~\ref{fig:AIAresponses}) is different from the others, the structures 
visible by each waveband are slightly different. The temperature response 
function for SDO/AIA 94 \AA~captures the top half of the temperatures obtained 
during the simulation, so that the heated magnetic structures  
inside the loop are visible in the intensity images, together with the average 
temperature inside the loop (Fig.~\ref{fig:AIA94}). The temperature response 
function of SDO/AIA 131 \AA~captures the middle 
of the temperature range during the simulation, so that the intensity images 
(Fig.~\ref{fig:AIA131}) shows mainly loop structure at the average temperature. 
The images obtained for SDO/AIA 211 \AA~(Fig.~\ref{fig:AIA211}) captures the 
maximum temperature during the simulation, with a contribution from the average 
and lower temperatures, as shown by its temperature response function (Fig.~\ref{fig:AIAresponses}).
The emission images of SDO/AIA 304 \AA~(Fig.~\ref{fig:AIAresponses}) are not presented 
in this study. This waveband captures the temperatures of the chromosphere and transition 
region, which are not included in our coronal loop model. Also, the emission images obtained for 
SDO/AIA 335 \AA~are not presented here. This temperature response function is broad and 
spans all the temperatures generated in the simulation. As a result, 70\% or more of 
the lower values of the intensity obtained by (Eq.3) have to be eliminated in 
order to produce intensity plots in which the loop structure is clearly visible. 

\section{Discussions and Conclusion}
\label{sec:concl}

\begin{figure*}
\centerline{
\includegraphics[width=15cm]{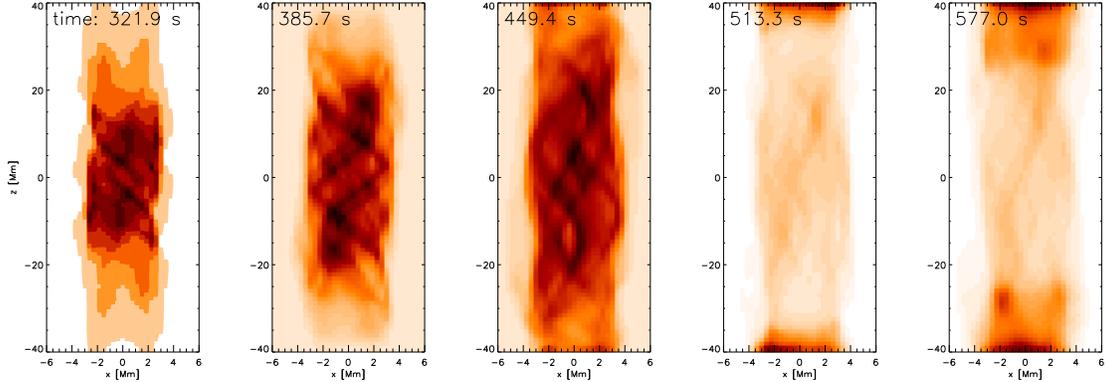}}
\caption{AIA 211 \AA~with parameters and times as in Figure \ref{fig:AIA94}. 
         The minimum (white) of the intensity scale is chosen so that the lowest 
         1\% of the values at time 321.9 s are eliminated from these plots. All 
         the exposures use the same reverse colour scale.  
         }
\label{fig:AIA211}
\end{figure*}

A highly twisted and kink unstable coronal loop is studied numerically by solving the resistive MHD equations for a 
fully ionised plasma and with the inclusion of the thermal conduction. The loop is 
initialised as a straight cylinder with a very high twist above the stability threshold leading to the 
kink instability. The simulations were initialised with physical parameters derived from the 
observations of a highly twisted coronal loop \citep{SrivEA10}. The initial magnetic field structure used in the simulations is 
considered as a kink-unstable 
force-free equilibrium where the twist varies with radius \citep{HoodEA09, BothaEA12}. 
Line of sight emission intensities were calculated from the simulation data using the 
corrected temperature response functions from  SDO/AIA. The different wavebands from the 
SDO/AIA exhibits different magnetic field structures, as each filter
temperature response function captures a different plasma temperature and related emissions. Therefore, it 
becomes plausible to study many high-resolution features of a highly twisted
kink-unstable coronal loop as observables of the various filters of SDO/AIA. 
During the evolution of the kink instability, the internal structure of the 
synthesized loop is shown in the generated images (Figs.~\ref{fig:AIA94}-\ref{fig:AIA211}). 
This structure is transformed into a simpler magnetic configuration as the kink instability triggers 
multiple reconnections that cause the straightening of the internal 
magnetic field. 
The kink instability heats the plasma near the current sheet due to the reconnection.  
This heat flux is transported along the magnetic field lines 
away from the reconnection sites due to thermal conduction. Therefore, the temperature maximum in the simulation is lower 
compared to the case when thermal conduction is excluded from the model \citep{BothaEA12}. 

The observations by \citet{SrivEA10} clearly shows that after the heating as well as activation of high twist
,the loop plasma is condensed down within few minutes and becomes invisible 
in coronal filters. The duration 
of the simulations exhibits 4.25 minutes of the nonlinear evolutionary phase of the 
kink instability. During this time span the average temperature of the loop increases 
(Fig.~\ref{fig:kink}, right-panel) while the maximum temperature with the thermal conduction is almost unchanged 
(Fig.~\ref{fig:kink}, left-panel). 
The reason may be that the current sheets evolve through the multiple reconnection 
process instead of continuous heating episode. Therefore, the heating location moves
in the simulated loop \citep{HoodEA09}.
Footpoint brightening due to compression of the plasma is also evident in the simulation 
, however,  it is not evident from the coronal observations by \cite{SrivEA10}. 
Although, \cite{SrivEA10} have observed brightpoints at chromospheric temperatures 
that may be the evidence of localized footpoint heating in lower atmosphere.  

In conclusion, we extend the works of \cite{BothaEA12} and show that how the kink
unstable observed loop will be synthesized in various high-temperature filters 
of SDO/AIA. The physical scenario and the technical details of the numerical simulation presented 
in the paper  are very similar to the Botha et al. (2012). However, the 
present work demonstrates about the synthesis of the observational 
signature of coronal kink instability in the various filters of SDO/AIA
and discuss its physical significance. More studies should be performed using forthcoming observations
from SDO/AIA regarding the helical kink unstable loops that may shed more light 
on the stringent simulation initial conditions, and can provide more understanding 
of the energy generation and transport in the coronal loop systems during 
the evolution of such instability.

\medskip
\noindent {\bf Acknowledgments}
We thank both the reviewers for their suggestions that improved the manuscript.
The use of LARE3d code is acknowledged in the present work. 
{\bf This work was partly funded by the UK Science and Technology Facilities Council under grant number ST/I000720/1.}
AKS acknowledges the financial support from DST-RFBR-P117 project, 
and also acknowledges Shobhna Srivastava for patient encouragements.
\medskip



\begin{thebibliography}{}

\bibitem[Arber et al.(1999)]{ArberEA99}
    Arber, T. D., Longbottom, A. W., \& Van der Linden, R. A. M. 
    1999, \apj, 517, 990

\bibitem[Arber et al.(2001)]{ArberEA01}
    Arber, T. D., Longbottom, A. W., Gerrard, C. L., \& Milne, A. M. 
    2001, J.\ Comput.\ Phys., 171, 151

\bibitem[Aschwanden \& Boerner(2011)]{AshBoe11}
    Aschwanden, M. J., \& Boerner, P., 2011, \apj, 732, 81


\bibitem[Botha et al.(2011)]{BothaEA11} 
    Botha, G. J. J., Arber, T. D., \& Hood, A. W. 
    2011, \aap, 525, A96

\bibitem[Botha et al.(2012)]{BothaEA12} Botha, G.~J.~J., Arber, 
T.~D., \& Srivastava, A.~K.\ 2012, \apj, 745, 53 


\bibitem[Baty \& Heyvaerts(1996)]{BatyHeyvaerts96}
    Baty, H., \& Heyvaerts, J. 1996, \aap, 308, 935 


\bibitem[Browning et al.(2008)]{BrowningEA2008}
    Browning, P. K., Gerrard, C., Hood, A. W., Kevis, R., \& 
    Van der Linden, R. A. M. 2008, \aap, 485, 837

\bibitem[Cho et al.(2009)]{Cho09} Cho, K.-S., Lee, J., Bong, 
S.-C., et al.\ 2009, \apj, 703, 1 


\bibitem[Foullon et al.(2011)]{Claire11} Foullon, C., Verwichte, 
E., Nakariakov, V.~M., Nykyri, K., \& Farrugia, C.~J.\ 2011, \apjl, 729, L8


\bibitem[Gerrard et al.(2002)]{GerrardEA02}
    Gerrard, C. L., Arber, T. D., \& Hood, A. W. 2002, \aap, 387, 687



\bibitem[Haynes \& Arber(2007)]{HaynesArber07} 
    Haynes, M., \& Arber, T. D. 2007, \aap, 467, 327

\bibitem[Haynes et 
al.(2008)]{Hy08} Haynes, M., Arber, T.~D., \& Verwichte, E.\ 2008, \aap, 479, 235 


\bibitem[Hood 
\& Priest(1979)]{Alan79} Hood, A.~W., \& Priest, E.~R.\ 1979, \solphys, 64, 303 


\bibitem[Hood et al.(2009)]{HoodEA09} 
    Hood, A. W., Browning, P. K., \& Van der Linden, R. A. M. 
    2009, \aap, 506, 913

\bibitem[Innes et 
al.(2012)]{Innes12} Innes, D.~E., Cameron, R.~H., Fletcher, L., Inhester, B., \& Solanki, S.~K.\ 2012, \aap, 540, L10 

\bibitem[Kliem \& T\"or\"ok (2006)]{Kleim06} Kliem, B., T\"or\"ok, T.\ 2006, Physical Review Letters, 96, 255002 

\bibitem[Kozlova \& Somov(2009)]{KozSom09}
    Kozlova, L. M., \& Somov, B. V. 
    2009, Moscow University Physics Bulletin, 64, 541

\bibitem[Kumar et al.(2010)]{Ku10} Kumar, P., Srivastava, 
A.~K., Somov, B.~V., et al.\ 2010, \apj, 723, 1651 


\bibitem[Liu et al.(2007)]{Liu07} Liu, R., Alexander, D., 
\& Gilbert, H.~R.\ 2007, \apj, 661, 1260 

\bibitem[Liu et al.(2008)]{Liu08} Liu, R., Gilbert, H.~R., 
Alexander, D., \& Su, Y.\ 2008, \apj, 680, 1508 

\bibitem[Miki\'c et al.(1990)]{MikicEA90} 
   Miki\'c, Z., Schnack, D. D., \& Van Hoven, G. 1990, \apj, 361, 690

\bibitem[Moser 
\& Bellan(2012)]{Moser12} Moser, A.~L., \& Bellan, P.~M.\ 2012, \nat, 482, 379 


\bibitem[O'Dwyer et al.(2010)]{ODwyerEA10} 
    O'Dwyer, B, Del Zanna, G., Mason, H. E., Weber, M. A., \& D. Tripathi, D.
    2010, \aap, 521, A21

\bibitem[Petrie \& Patrikeeva(2009)]{PetPat09}
    Petrie, G. J. D., \& Patrikeeva, I. 2009, \apj, 699, 871

\bibitem[Priest(2000)]{Priest00}
    Priest, E. R. 2000, Solar Magnetohydrodynamics, 
    Geophysics and Astrophysics Monographs, Volume 21,  
    (Dordrecht: D. Reidel Publishing Company), Section 2.3.2
    

\bibitem[Spitzer-H\"arm(1953)]{SpitzerHarm53}
    Spitzer, L., \& H\"arm, R. 1953, Physical Review, 89, 977

\bibitem[Srivastava et al.(2010)]{SrivEA10} 
    Srivastava, A. K., Zaqarashvili, T. V., Kumar, P., \& Khodachenko, M. L. 
    2010, \apj, 715, 292



\bibitem[Young et al.(2009)]{YoungEA09} 
    Young, P. R., Watanabe, T., Hara, H., \& Mariska, J. T. 
    2009, \aap, 495, 587

\bibitem[Zaqarashvili et 
al.(2010)]{Zaqa10} Zaqarashvili, T.~V., D{\'{\i}}az, A.~J., Oliver, R., \& Ballester, J.~L.\ 2010, \aap, 516, A84 


\end{thebibliography}
\end{document}